\documentclass[12pt]{article}

\usepackage{graphicx}
\graphicspath{{./figures/}}
\usepackage{dcolumn}
\usepackage{bm}
\usepackage{amsfonts}
\usepackage{amsthm}
\usepackage{amsmath}
\usepackage{amssymb}
\usepackage{color}
\usepackage{textcomp}
\usepackage{subfig}

\usepackage{hyperref}

\usepackage[utf8]{inputenc}
\usepackage{lipsum}
\usepackage{centernot}
\usepackage{feynmp}

\def\be{\begin{equation}}
\def\ee{\end{equation}}
\def\ba{\begin{eqnarray}}
\def\ea{\end{eqnarray}}

\def\bdm{\begin{displaymath}}
\def\edm{\end{displaymath}}

\def\bq{\begin{quote}}
\def\eq{\end{quote}}

 at 10truept

\def\beastar{\begin{eqnarray*}}
\def\eeastar{\end{eqnarray*}}       

\newcommand{\bea}{\begin{eqnarray}}
\newcommand{\eea}{\end{eqnarray}}

\newcommand{\bi}{\begin{itemize}}
\newcommand{\ei}{\end{itemize}}

\newcommand{\beq}{\begin{equation}}
\newcommand{\eeq}{\end{equation}}
\newcommand{\beqa}{\begin{eqnarray}}
\newcommand{\eeqa}{\end{eqnarray}}

\def\ltap{\ \raise.3ex\hbox{$<$\kern-.75em\lower1ex\hbox{$\sim$}}\ }
\def\gtap{\ \raise.3ex\hbox{$>$\kern-.75em\lower1ex\hbox{$\sim$}}\ }
\def\gl{\ \raise.5ex\hbox{$>$}\kern-.8em\lower.5ex\hbox{$<$}\ }
\def\roughly#1{\raise.3ex\hbox{$#1$\kern-.75em\lower1ex\hbox{$\sim$}}}

\begin{document}

\thispagestyle{empty}
\begin{flushright}
June 2016
\end{flushright}
\vspace*{1.25cm}
\begin{center}
{\Large \bf Vacuum Energy Sequestering and Graviton Loops}\\

\vspace*{1.6cm} {\large Nemanja Kaloper$^{a, }$\footnote{\tt
kaloper@physics.ucdavis.edu} and Antonio Padilla$^{b, }$\footnote{\tt
antonio.padilla@nottingham.ac.uk} }\\
\vspace{.5cm} {\em $^a$Department of Physics, University of
California, Davis, CA 95616, USA}\\
\vspace{.5cm} {\em $^b$School of Physics and Astronomy, 
University of Nottingham, Nottingham NG7 2RD, UK}\\

\vspace{1.5cm} ABSTRACT
\end{center}
We recently formulated a local mechanism of vacuum energy sequester. This mechanism automatically removes all matter loop contributions to vacuum energy from the stress energy tensor which sources the curvature. Here we adapt the local vacuum energy sequestering mechanism to 
also cancel all the vacuum energy loops involving virtual gravitons, in addition to the vacuum energy generated by matter fields alone.

\vfill \setcounter{page}{0} \setcounter{footnote}{0}
\newpage

The cosmological constant problem \cite{zeldovich,wilczek,wein, pol,cliff,me} follows from two simple statements:  in a local Quantum Field Theory (QFT), off-shell dynamics facilitated by virtual particles renormalizes the Lagrangian, and in General Relativity (GR), the universality of gravitational couplings ensures that all energy gravitates  in the same way. The vacuum energy induced by virtual particles, scaling as $({\rm cutoff})^4$,  behaves just like a cosmological constant, curving the geometry of space-time even in vacuum. Cosmological observations constrain the vacuum energy density to be less than $(\textrm{meV})^4$. This is at least 60 orders of magnitude below a naive theoretical estimate based on the possible value of the cutoff of the low energy QFT and the absence of a dynamical cancellation mechanism below that cutoff.  

In QFT coupled to GR, the real cosmological constant problem is more subtle. Consider a low energy effective field theory (EFT), defined up to a cutoff $M$. In the absence of symmetries that can enforce
cancellations\footnote{For example, supersymmetry and/or exact scale invariance}, its vacuum energy loops generically scale as $M^4$ \cite{wein,akhmedov,sirlin}. This means that the observable that is corrected by the vacuum energy loops -- the curvature of the empty space -- is UV sensitive, and must be renormalized.  Its physical value is the sum total of the quantum vacuum contributions and the bare counterterm.  This is a finite quantity which cannot be predicted, but must be measured. When the cutoff-dependent pieces are subtracted, the renormalized cosmological constant is a function of the finite contributions coming from all the physical scales $m$ below the cutoff and the arbitrary subtraction scale $\bar M$, where it is measured. 
The problem is that the renormalized cosmological constant depends on the {\it powers} of the physical scales below the cutoff, implying that it can be greatly affected by any existing heavy field beyond the threshold of  local measurements at sub-TeV scales.  

This sensitivity to the scales that govern unknown physics is {\it the} problem. To reproduce the measured value at any level of perturbation theory we must choose the bare counterterm with great precision in the units of the cutoff. This counterterm serves as an avatar for the unknown high scales, and its choice is not robust against modifications of the UV sector of the theory, even though the physics which sets the value of the renormalized cosmological constant is in the far infra-red. If we view the bare counterterm as a cosmological initial condition,  this shows a great deal of sensitivity to the cosmological initial conditions in the vacuum energy sector. 

These difficulties with the cosmological constant should be contrasted with the mass of the electron in QED. The renormalized electron mass depends {\it logarithmically} on the cutoff due to the restoration of the electron chiral symmetry in the massless limit \cite{weisskopf,chiral}. This means that the electron mass cannot be calculated in QFT, but must also be measured. The important point, however, is that  the UV sensitivity is only logarithmic, too mild to significantly affect the cancellation between the quantum corrections and the counterterm. The quantum corrections are proportional to the symmetry-breaking parameters -- the electron mass -- so  they are never much larger than the renormalized value at any loop order. This is how supersymmetry and/or scaling symmetry could stabilize the vacuum energy in principle. The problem is that if these symmetries exist, the relevant numerical scales controlling their  breaking are much larger than the values required by observation. Nevertheless, this argument has been the {\it raison d'\^{e}tre} behind the quest for the `dynamical solution' of the cosmological constant's radiative instability, obstructed by Weinberg's venerable no-go theorem \cite{wein}.

Following early work \cite{andrei,tseytlin,self,degrav}, we have recently proposed a mechanism for eliminating the contributions to vacuum energy from matter loops, dubbed {\it vacuum energy sequestering}  \cite{KP1,KPSZ,KPS}.  The procedure is a very conservative, minimalistic  modification of the gravitational sector of the theory, deviating from the on-shell behavior of GR only at the global level, in the infinite wavelength limit.  Exploiting the fact that vacuum energy is the only source of curvature that is spacetime filling (and locally constant), we introduce constraints that operate at the largest wavelengths,  along with new gauge symmetries that prohibit any additional local degrees of freedom. The constraints ensure that counterterms always cancel the power-sensitive contributions to vacuum energy from matter loops,  so that they do not gravitate.  Gravity is sourced by a renormalized vacuum energy that is radiatively stable, albeit with a value that is incalculable and should be set by measurement (just like the electron mass). For additional recent explorations see \cite{kluson,oda}.

Here we pursue a modification of the mechanism that addresses vacuum energy loops that include virtual gravitons. We show that higher dimensional operators can be used as conjugate variables to construct constraints that sequester vacuum energy loops with graviton lines from the gravitational field equations. The mechanism can be embedded in the formulation of gravity as a quantum EFT defined up to some scale $M \lesssim  M_{Pl}$, below which the EFT is unitary and weakly coupled, following the ``classic lore" of \cite{tHV, GS}, along with more recent ideas developed by Donoghue \cite{Don}. 

To set the stage, we review the local matter vacuum energy sequestering, given by \cite{KPSZ}
\be\label{localaction1}
S = \int d^4 x \left\{ \sqrt{g} \left[ \frac{\kappa^{2}(x)}{2} R  - \Lambda(x) - {\cal L}_m( g^{\mu\nu} , \Phi) \right] +  \frac{\epsilon^{\mu\nu\lambda\sigma}}{4!} 
\left[\sigma\left(\frac{ \Lambda}{ \mu^4}\right){F_{\mu\nu\lambda\sigma} } +\hat \sigma\left(\frac{ \kappa^{2}}{ M_{Pl}^2}\right){\hat{F}_{\mu\nu\lambda\sigma}} \right] \right\} \, .
\ee
In addition to the metric and the matter fields $\Phi$ we include a pair of $4$-forms $F_{\mu\nu\lambda\sigma} = 4\partial_{[\mu}A_{\nu\lambda\sigma]}$ and $\hat{F}_{\mu\nu\lambda\sigma} = 4\partial_{[\mu}\hat{A}_{\nu\lambda\sigma]}$ and a pair of scalar fields $\kappa(x)$ and $\Lambda(x)$. The gauge symmetries of the $4$-forms enforce that these scalars  have no fluctuating modes \cite{henteitelboim}. The arguments of the smooth functions $\sigma$ and $\hat{\sigma}$ are normalized to the two high energy scales $\mu$, $M_{Pl}$ which are close to the EFT cutoff, $M \lesssim \mu, M_{Pl}$. The properties of $\sigma,\hat \sigma$ are discussed in \cite{KPS}; we stress that they cannot be linear functions to avoid any hidden fine tunings in the vacuum energy cancellations. The matter sector ${\cal L}_m$ couples to the metric minimally. The last two terms of (\ref{localaction1}) are a non-gravitating, topological sector by virtue of the absence of the metric. 

To get the dynamics, we could now vary (\ref{localaction1}), obtain the local field equations, and then concentrate on the global, infinite wavelength sector to address the curvature (in)sensitivity to vacuum energy loops \cite{KPSZ}. That yields very similar equations to the original global constraints of \cite{KP1}, with the main difference being that the global constraints appear as integrals of the local field equations, zooming in on the vacuum energy \cite{degrav,KPSZ}. In particular, the cancellation works in spacetimes with infinite 4-volume and finite field theory scales. 

Since we are mainly interested in the mechanics of cancellation of vacuum energy loops, we can shortcut our analysis by working ``in the action" and focusing on the global sector from the start. To this end, we integrate out the $3$-forms from (\ref{localaction1}), bearing in mind that they impose the `rigidity' of the scalars $\kappa$ and $\Lambda$, forcing them to be spacetime constants.  The effective dynamics of the global sector of the theory is now controlled by
\be\label{globalaction1}
S= \int d^4 x \sqrt{g} \left[ \frac{\kappa^{2}}{2} R  - \Lambda - {\cal L}_m( g^{\mu\nu} , \Phi) \right]  +
\sigma\left(\frac{ \Lambda}{ \mu^4}\right)c+\hat \sigma\left(\frac{ \kappa^{2}}{ M_{Pl}^2}\right)\hat c \, .
\ee
Here $c$ and $\hat c$ respectively describe the flux of $3$-forms, $A$ and $\hat A$, with the additional constraints obtained by varying with respect to the {\it rigid} scalars $\kappa^2$ and $\Lambda$. The field equations are
\be 
\kappa^{2} G^\mu{}_\nu = T^\mu{}_\nu-\Lambda \delta^\mu{}_\nu \, , ~~~~~~
 \frac{\sigma'}{\mu^4} c = \int \sqrt{g} d^4 x, ~~~~~~~
\frac{\hat \sigma'}{M_{Pl}^2}\hat c =-\frac{1}{2} \int R \sqrt{g} d^4 x \, . \label{glob1}
\ee 
Tracing the gravity equation and averaging over spacetime fixes $\Lambda$ in terms of $\langle T^\alpha{}_\alpha \rangle$ and $\langle R \rangle$, where $T^\alpha{}_\alpha = g^{\mu\nu} T_{\mu\nu}$ is the {\it regularized} trace of the matter stress energy tensor at a given loop order and angled brackets denote the spacetime average.  This gives $\Lambda = \langle T^\alpha{}_\alpha \rangle/4 + \Delta \Lambda$ where 
$\Delta \Lambda= \kappa^{2}  \langle R \rangle /4=-\frac{\mu^4}{2} \frac{\kappa^{2}  \hat \sigma'}{M_{Pl}^2 \sigma'}\frac{\hat c}{c} $, the last equality
following from the ratio of the  global equations in (\ref{glob1}). Inserting this expression into the gravity equation yields
\be \label{effGeq0}
\kappa^{2} G^\mu{}_\nu = T^\mu{}_\nu - \frac{1}{4} \delta^\mu{}_\nu \langle T^\alpha{}_\alpha \rangle - \Delta \Lambda  \delta^\mu{}_\nu \, .
\ee
This equation shows that vacuum energy $ \langle vac |T^\mu{}_\nu| vac \rangle  = -\delta^\mu{}_\nu V_{vac}$ completely drops out of the gravitational dynamics. Matter radiative corrections do affect the finite renormalized cosmological constant $\Delta \Lambda$ in (\ref{effGeq0}) since $\Lambda \to \Lambda+{\cal O}( M^4)$, and $\kappa^2 \to \kappa^2+{\cal O}(M^2)$ \cite{Demers}. However, when $\sigma$ and $\hat \sigma$ are smooth (ie $\sigma({\cal O}(1) z ) \sim {\cal O}(1)\sigma(z)$, etc) and non-degenerate functions \cite{KPSZ,KPS}, the renormalized cosmological constant 
is radiatively stable since $M \lesssim \mu, M_{Pl}$ .

The vacuum energy cancellation described above is enforced by two approximate symmetries of the theory  \cite{KP1}. The first is the shift symmetry ${\cal L}_m \to {\cal L}_m+\nu^4$, $\Lambda\to \Lambda-\nu^4$, where $\nu$ is a constant, which is broken by the topological terms, yet restored in the limit $c/\mu^4 \to 0$.  The second approximate symmetry is the scaling symmetry of \cite{KP1}, which in terms of the  variables we employ here dwells in the $\kappa^2$ sector. To see it consider metric and $\kappa^2$ fluctuations about a flat background in the limit of vanishing cosmological constant. Setting  $\kappa^2=M_{Pl}^2(1+ {\phi}/{M_{Pl}})$ and $g_{\mu\nu}=\eta_{\mu\nu}+{h_{\mu\nu}}/{M_{Pl}}$, the theory is invariant under $\phi \to  \phi+\hat \nu$ in the limit $\hat c/M_{Pl} \to 0$, $M_{Pl} \to \infty$. Clearly this symmetry is broken at finite $M_{Pl}$. While the breaking is weak for the matter sector loops, it implies that graviton loops will not be cancelled in the theory (\ref{localaction1}).

To see this explicitly consider the vacuum energy renormalization of the 1PI effective action by loops that involve both matter and gravitons. 
For simplicity, we compute them in a locally Lorentzian frame, treating the background geometry as flat. This correctly captures all the UV contributions. 
Expanding in the gravitational coupling, the result is
\be
-\left[ a_0 M^4 +a_1 \frac{M^6}{\kappa^2}  +a_2\frac{ M^8}{\kappa^4}+\ldots \right] \int  \sqrt{g} d^4 x \, ,
\ee
where $a_i \sim {\cal O}(1)$. The terms $\sim M^4$ are the contributions from matter vacuum energy loops. Pure gravity loop diagrams also contribute in the same way. They are automatically sequestered away from sourcing curvature in the theory (\ref{localaction1}). The terms  that go as powers of $1/\kappa^2$  contain graviton interactions. The $\sim M^6/\kappa^2$ terms can arise from diagrams\footnote{Here we also include the pure gravity loops, although eg. they vanish around flat space in dimensional regularization since they are given by scaleless integrals \cite{tHV,akhmedov,sirlin}. However, in principle they are still sensitive to the details of the UV and IR regulator, as is seen from mixed matter-gravity loops involving heavy fields and/or in curved backgrounds.} such as

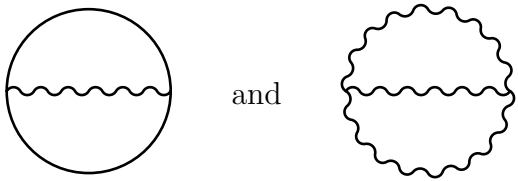
\begin{equation}
\unitlength=0.5mm
\begin{fmffile}{grav2}
%
\begin{gathered}
 \begin{fmfgraph}(75,75)
    \fmfleft{i}
\fmfright{o}
\fmf{phantom,tension=5}{i,v1}
\fmf{phantom,tension=5}{v2,o}
\fmf{plain,left,tension=0.4}{v1,v2,v1}
\fmf{photon}{v1,v2}
  \end{fmfgraph}
  \end{gathered}
\end{fmffile}
\textrm{and}
\begin{fmffile}{grav4}
%
 \begin{gathered}
 \begin{fmfgraph}(75,75)
    \fmfleft{i}
\fmfright{o}
\fmf{phantom,tension=5}{i,v1}
\fmf{phantom,tension=5}{v2,o}
\fmf{photon,left,tension=0.4}{v1,v2,v1}
\fmf{photon}{v1,v2}
  \end{fmfgraph}
\end{gathered}
\end{fmffile}
\nonumber
\end{equation}

\noindent where the solid lines denote matter propagators, and the wiggly lines  are gravitons.   For $\kappa \sim M_{Pl}$, and a cutoff as low as $M \sim$ TeV, the numerical value of their regularized contributions  is already thirty orders of magnitude above the dark energy scale. 
The contributions $\sim M^8/\kappa^4$ come from diagrams with either more, or higher order, graviton interactions, such as

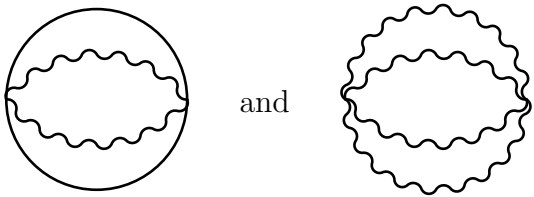
\be
\unitlength=0.5mm
\begin{fmffile}{grav3}
%
\begin{gathered} 
 \begin{fmfgraph}(75,75)
    \fmfleft{i}
\fmfright{o}
\fmf{phantom,tension=10}{i,v1}
\fmf{phantom,tension=10}{v2,o}
\fmf{plain,left,tension=0.4}{v1,v2,v1}
\fmf{photon,left=0.5}{v1,v2}
\fmf{photon,right=0.5}{v1,v2}
  \end{fmfgraph}
\end{gathered}
\end{fmffile}
\textrm{and} 
\unitlength=0.5mm
\begin{fmffile}{grav5}
%
\begin{gathered}
 \begin{fmfgraph}(75,75)
    \fmfleft{i}
\fmfright{o}
\fmf{phantom,tension=10}{i,v1}
\fmf{phantom,tension=10}{v2,o}
\fmf{photon,left,tension=0.4}{v1,v2,v1}
\fmf{photon,left=0.5}{v1,v2}
\fmf{photon,right=0.5}{v1,v2}
  \end{fmfgraph}
 \end{gathered} 
\end{fmffile}\nonumber
\ee

\noindent   Curiously, if the cutoff is as low as TeV their regularized values would be at the dark energy scale or below. However, for higher cutoffs they are dangerously large.

In any case, the $\kappa^2$-dependent corrections to vacuum energy will not be sequestered in the theory (\ref{localaction1}). They scale differently with the cutoff, spoiling the cancellation implied by  (the trace of)  Einstein equations and  the geometric constraint found by varying  (\ref{globalaction1}) with respect to  the rigid scalar $\kappa$. Indeed, calculating the vacuum field equations we find that
\ba
&&~~~~~~~~~~~~~~~~ \kappa^{2} G^\mu{}_\nu = -\left(\Lambda + a_0 M^4 +a_1 \frac{M^6}{\kappa^2}  +a_2\frac{ M^8}{\kappa^4}+\ldots \right)\delta^\mu{}_\nu \, , \label{graveqa} \nonumber \\
 &&\frac{\sigma'}{\mu^4} c = \int \sqrt{g} d^4 x, ~~~~~~~~~~
\frac{\hat \sigma'}{M_{Pl}^2}\hat c =-\frac{1}{2} \int \left(R +2a_1 \frac{M^6}{\kappa^4}  +4a_2\frac{ M^8}{\kappa^6}+\ldots \right) \sqrt{g} d^4 x \, , \label{glob2} 
\ea
Tracing, averaging over the spacetime, and eliminating constraints yields
\be
\kappa^{2} G^\mu{}_\nu =  - \left[\Delta \Lambda -a_1 \frac{M^6}{2\kappa^2}  -a_2\frac{ M^8}{\kappa^4}+\ldots  \right]  \delta^\mu{}_\nu \, .
\ee
with 
$\Delta \Lambda= \kappa^{2}  \langle R \rangle /4=-\frac{\mu^4}{2} \frac{\kappa^{2}  \hat \sigma'}{M_{Pl}^2 \sigma'}\frac{\hat c}{c} $ as before. 
The regularized vacuum energy terms independent of $\kappa$ cancel out, and the only residual dependence remains through the radiatively stable finite term $\Delta \Lambda$. In contrast, the $\kappa$ dependent pieces do not  cancel. Clearly, the largest contributions come from terms $\sim  \frac{M^6}{2\kappa^2}$, but others are in principle dangerous too, requiring some additional mechanism to keep them under control. 

The main purpose of this Letter is to point out that such a mechanism can be obtained by a straightforward modification of the local sequestering theory (\ref{localaction1}). The logic behind local sequestering was threefold: (i) promote the gravitational parameters $\kappa^2, \Lambda$ into local fields;  (ii) project out their local fluctuations by the gauge symmetries of the 4-forms;  (iii) retain variational equations with respect to the rigid fields $\kappa^2,\Lambda$  because they fix the counterterms and divert radiative instabilities away from the metric and into the physically unobservable sector of local 4-form fluctuations. The key condition arises from the variation with respect to $\kappa$, whose global limit is the condition that the spacetime average of the scalar curvature is fixed by a radiatively stable quantity $\Delta \Lambda$, controlled by the ratio of the 4-form fluxes. 
By Einstein's equations,  the cutoff-dominated terms in $\Lambda - \langle T^\alpha{}_\alpha \rangle/4$ automatically cancel.

One can immediately verify that a qualitatively similar condition would follow from vanishing of the spacetime average of any generic curvature invariant not constructed purely out of the Weyl tensor and the traceless part of the Ricci tensor. Any such invariant would not be scale invariant and would therefore involve the Ricci scalar.  By the vacuum Einstein's equation, it follows that this curvature invariant  would be polynomial in the difference $\Lambda - \langle T^\alpha{}_\alpha \rangle/4$. Fixing it on shell by a variational principle to a radiatively stable quantity would yield a behavior similar to that which follows from (\ref{localaction1}).  Furthermore, if the variational constraint is not directly dependent on the Planck scale, the cancellation of Planck-mass dependent vacuum energy contributions -- namely, those involving graviton virtual lines -- will also cancel from the residual stress energy tensor to leading order. 

Although we can construct many suitable curvature invariants in four dimensions, one immediately arises as a most minimal candidate: 
the Gauss-Bonnet invariant, $R_{GB}=R_{\mu\nu\alpha\beta}^2-4R_{\mu\nu}^2+R^2$. Because it is a total derivative, adding it to the action merely changes the topological sector, and does not affect any local phenomena at finite wavelength. Since it involves the Ricci scalar, and so is not scale invariant, it yields a desired constraint that picks the correct counterterms to sequester all large contributions from loops from the source of the Einstein's equations. 

Let us demonstrate this explicitly. We start with the action
\be\label{localaction2}
S= \int d^4 x \left\{\sqrt{g} \left[ \frac{M_{Pl}^2}{2} R+\theta(x)R_{GB}  - \Lambda(x) - {\cal L}_m 
\right] +  
\frac{\epsilon^{\mu\nu\lambda\sigma}}{4!} \left[\sigma\left(\frac{ \Lambda}{ \mu^4}\right){F_{\mu\nu\lambda\sigma} } +\hat \sigma\left(\theta\right){\hat{F}_{\mu\nu\lambda\sigma}} \right] \right\}\, .
\ee
where we now vary over the  auxiliary scalar $\theta(x)$ controlling the Gauss-Bonnet coupling. Again for simplicity, we focus only
on the global limit of the theory, and integrate out the $3$-forms as before, yielding an effective action
\be\label{globalaction2}
S= \int d^4 x \sqrt{g} \left[ \frac{M_{Pl}^2}{2} R+\theta R_{GB}  - \Lambda - {\cal L}_m( g^{\mu\nu} , \Phi) \right] +
\sigma\left(\frac{ \Lambda}{ \mu^4}\right)c+\hat \sigma\left(\theta \right)\hat c \, .
\ee
The two scalars are now rigid, with no local variations off-shell, and $c$ and $\hat c$ are the fluxes of the $3$-forms through the boundary.  The resulting field equations are
\be
M_{Pl}^{2} G^\mu{}_\nu = T^\mu{}_\nu-\Lambda \delta^\mu{}_\nu \, , ~~~~~~
 \frac{\sigma'}{\mu^4} c = \int \sqrt{g} d^4 x, ~~~~~~
\hat \sigma' \hat c =- \int R_{GB} \sqrt{g} d^4 x \, . \label{glob} 
\ee
We can write $R_{GB}=W_{\mu\nu\alpha\beta}^2-2\left(R_{\mu\nu}-\frac14 R g_{\mu\nu}\right)^2+\frac16 R^2$, where $W_{\mu\nu\alpha\beta}$ is the Weyl tensor, whose contribution drops out in the vacuum.  Again, taking traces, averages, and eliminating constraints, we find
\be \label{effGeq}
M_{Pl}^{2} G^\mu{}_\nu = T^\mu{}_\nu - \frac{1}{4} \delta^\mu{}_\nu \langle T^\alpha{}_\alpha \rangle - \Delta \Lambda  \delta^\mu{}_\nu \, ,
\ee
where now $\Delta \Lambda$ satisfies 
\be
\Delta \Lambda^2=\frac{3 M_{Pl}^4 }{8} \left[\langle R_{GB} \rangle-\langle W_{\mu\nu\alpha\beta}^2 \rangle +\frac{2}{M_{Pl}^4} \langle (T_{\mu\nu}-\frac14 T  g_{\mu\nu})^2\rangle \right. 
\left. -\frac{1}{6 M_{Pl}^4}\left(\langle T^2 \rangle-\langle T\rangle^2 \right)\right] \, .
\ee
The spacetime average of the Gauss-Bonnet invariant is constrained by the ratio of the global equations, such that 
\be
\langle R_{GB} \rangle=-\mu^4\frac{ \hat \sigma'}{ \sigma'}\frac{\hat c}{c} \, .
\ee
Of course, we stress that the full system of equations from (\ref{localaction2}) is more complicated. However the global limit is faithfully reproduced by the equations  obtained from (\ref{globalaction2}).

As before, the regularized vacuum energy,  $\langle vac |T^\mu{}_\nu | vac \rangle  = -\delta^\mu{}_\nu V_{vac}$, completely drops out of the first two terms in equation (\ref{effGeq}). By its scale invariance, the Weyl tensor contribution vanishes, and so radiative corrections  can only affect $\Delta \Lambda$ through its dependence on $\langle R_{GB} \rangle$. These corrections yield $\Lambda \to \Lambda +{\cal O}(M^4)$,  and $\theta \to \theta +{\cal O} (1)\ln({M}/{m})$, where $m$ is a typical mass scale in the EFT \cite{Demers}. Therefore, by the same line of reasoning given after Eq. (\ref{effGeq0}), we see that the corresponding source of curvature is radiatively stable.  This cancellation now occurs for  both the matter vacuum energy contributions, and for the vacuum energy loops involving virtual gravitons. The fact that Gauss-Bonnet is a topological invariant ensures that graviton loops cannot introduce any additional $\theta$ dependence in the off-shell effective action. One might worry about generating extra dependence on the rigid scalars from background curvature effects and the IR corrections they induce, which have so far been neglected. However these corrections are suppressed by the background curvature scale, and one expects them to be harmless.  Other radiatively induced curvature corrections, obtained by renormalizing (\ref{localaction2}) will likewise remain subleading below the cutoff $M \lesssim M_{Pl}$. 

The reason behind the improved behavior of the global sector (\ref{globalaction2}) of the theory (\ref{localaction2})  over the global sector of the theory (\ref{localaction1}) is that the second approximate shift symmetry now involves $\theta \rightarrow \theta + \alpha$, which is only broken by the topological terms in the theory. The bulk terms remain invariant even at finite $M_{Pl}$.  This means that as $\hat c \to 0$, but keeping $M_{Pl}$ finite, the symmetry is completely restored. This improved approximate shift symmetry, which was absent in our previous set-up at finite $M_{Pl}$, yields the cancellation of the cutoff dominated vacuum energy contributions that include virtual gravitons. The fact that the symmetry is broken only by the topological terms  prevents the generation of potential terms in $\theta$ that would otherwise spoil the vacuum energy sequester.

To summarize, in this Letter we have adapted the vacuum energy sequestering mechanism to yield a cancellation of all cutoff-dominated vacuum energy contributions, computed in the loop expansion using arbitrary bubble diagrams. We include all such diagrams, with or without virtual gravitons, treating gravity as an effective field theory with a cutoff below the Planck scale.  The generalization utilizes the Gauss-Bonnet topological invariant, yielding a better approximate shift symmetry that remains unbroken in the bulk even at finite $M_{Pl}$.  This generalization should be viewed as a particular example of what may well be a considerably broader class of models. The guideline may be that in formulating the low energy effects of quantum gravity, one may need to promote   all the UV sensitive `couplings' in the theory to independent fields, which are ``stiffened" by their mixing with the non-gravitating 4-form sectors. The gauge symmetries of the 4-forms render the local fluctuations of the new fields unphysical. In turn the rigid fields divert the vacuum energy contributions dominated by the cutoff into the local part of the 4-form fields, which do not gravitate. 

In our original set-up this logic applied to the bare cosmological constant and the Planck mass. Here we used the bare cosmological constant and the Gauss-Bonnet coupling. We can imagine combining the two frameworks into one, by writing the action 
\ba\label{localaction3}
S &=& \int d^4 x \sqrt{g} \left[ \frac{\kappa^{2}(x)}{2} R  - \Lambda(x) - {\cal L}_m( g^{\mu\nu} , \Phi) +\theta(x) R_{GB} +\ldots \right] \nonumber \\
&&  +\int \frac{dx^{\mu}dx^{\nu} \ldots }{4!} 
\left[\sigma_1\left(\frac{ \Lambda}{ \mu^4}\right){F^{(1)}_{\mu\nu\lambda\sigma} }+\sigma_2\left(\frac{ \kappa^{2}}{ M_{Pl}^2}\right){{F}^{(2)}_{\mu\nu\lambda\sigma}}  +\sigma_3\left(\theta \right){F^{(3)}_{\mu\nu\lambda\sigma} } +\ldots  ~~~ \right] \, 
\ea
Here instead of two constraint equations we would have three of them. For non-degenerate functions $\sigma_i$ there would be no fine tunings in the theory, since the three constraints would fix three independent quantities $\kappa^2, \Lambda, \theta$ in terms of the three 4-form fluxes.
Vacuum loops of both matter and gravitons should still be sequestered in this theory, as long as all the $\sigma$'s satisfy the generic smoothness condition, $\sigma\left({\cal O}(1)z\right)\sim {\cal O}(1)\sigma(z)$. The reason is that loop corrections are still guaranteed to be independent of $\theta$, preserving the effectiveness of the geometric constraint arising from $\theta$ variation.  From a symmetry perspective, we see that in the limit of vanishing 4-form flux, the action is invariant under constant shifts in $\theta$, even at finite values of the other couplings. This symmetry enhancement protects the observable part of the cosmological constant from large corrections. We note that the theory (\ref{localaction3}) shares some features with the frameworks introduced in the past for a phenomenologically motivated attempt to resolve cosmological singularities \cite{markov}.
We believe that it would be interesting to explore cosmological and phenomenological properties of this and similar theories. 

\vskip.5cm

{\bf Acknowledgments}: 
We would like to thank T. Hamill, D. Stefanyszyn and G. Zahariade for useful discussions. N.K. is supported in part by the DOE Grant DE-SC0009999. AP was funded by a Royal Society URF.

\end{document}